\documentstyle[aps,multicol,epsf,prb]{revtex}   
\begin{document}
\draft
\title{Critical current from dynamical boundary instability for fully frustrated Josephson junction arrays}
\author {Beom Jun Kim and  Petter Minnhagen}
\address {Department of Theoretical Physics,
Ume{\aa} University,
901 87 Ume{\aa}, Sweden}
\preprint{\today}
\maketitle
\begin{abstract}

We investigate numerically the critical current of two-dimensional fully
frustrated arrays of resistively shunted Josephson junctions at zero
temperature.  It is shown that a domino-type mechanism is responsible for the
existence of a critical current lower than the one predicted from the
translationally invariant flux lattice. This domino mechanism is demonstrated
for uniform-current injection as well as for various busbar conditions.  It is
also found that inhomogeneities close to the contacts makes it harder for the
domino propagation to start, which increases the critical current towards the
value based on the translational invariance.  This domino-type vortex motion
can be observed in experiments as  voltage pulses propagating from the contacts
through the array.

\end{abstract}


\begin{multicols}{2}
The two-dimensional (2D) Josephson junction arrays  (JJA's) are
ideal testing grounds for static and dynamic properties related to
vortices and constitute an intriguing field of
physics by itself.~\cite{newrock} They can be fabricated with high
precision and are ideally suited for experiments and simulation studies, as well as
theoretical approaches.~\cite{newrock}
Vortex physics is also of great importance for understanding the
properties of high-$T_c$ superconductors, in particular, the interplay between
vortex correlations, pinning, and dynamics.~\cite{blatter} The present
paper describes an interesting  example of such an interplay.  
It focuses on what happens at the current
threshold for the onset of resistance in the case of a 2D JJA
in a perpendicular magnetic field. The threshold is in such a case
caused by the onset of motion of the vortices induced by the external
magnetic field. We find that this onset of vortex motion is caused
by a domino-type effect corresponding to voltage pulses
traveling through the sample.

The value of the critical currents in 2D JJA's have been investigated theoretically by several 
authors.~\cite{teitel}
These studies have revealed that the zero-temperature critical current $I_c$ depends
strongly on the value of the frustration $f$ defined by the number of flux quanta
per plaquette induced by the external magnetic field, and they unanimously have obtained the value $I_c = \sqrt{2} - 1
\approx 0.41$ (in units of the critical current of a single junction) for the 
fully frustrated ($f=1/2$) square array.
On the other hand, numerical calculations 
based on the resistively shunted junction (RSJ) model with the uniform-current injection 
method (see Ref.~\onlinecite{khlee} for details) have persistently given the value 
$I_c = 0.35(1)$.~\cite{khlee,free,octavio,simkin98,bjkim-s}
The discrepancy has usually been attributed to either finite-size effects
or an artifact of the uniform-current injection method, but without
any really substantiating evidence for either claim. 
As to the first suggestion, the same numerical value has now been
found in a very large array 
($256\times 256$) (Ref.~\onlinecite{bjkim-s}) and 
the absence of any size dependence up to this array size appears to preclude the
possibility of an explanation in terms of finite-size effects. 
The second possibility also by now seems less likely  since
attempts to model the injection of current through 
busbars, in close analogy with actual experimental situations (see
Ref.~\onlinecite{simkin97} for details),
in fact seem to give an even smaller value.~\cite{free,simkin98,bjkim-s}
Furthermore, in Ref.~\onlinecite{simkin98} an improved busbar method was proposed
and $I_c\approx 0.35$, close to the one obtained for the uniform-current injection, was
found. Thus a value lower than the theoretically
predicted one has been obtained with a variety of injection methods.

In the present paper we resolve this long-standing discrepancy
and find that the true critical current is in
fact lower than the theoretically predicted one and corresponds to a
domino-type mechanism starting from the contacts
which breaks the translational invariance of the vortex
lattice. This means that the theoretically predicted value, which presumes
translational invariance, in fact gives an upper limit of $I_c$.
We conclude that
the true critical current is always lower and that the precise value,
because it is given by a domino mechanism, is sensitive to the inhomogeneities 
close to the contacts.

\begin{figure}
\centerline{\epsfxsize=8cm \epsfbox{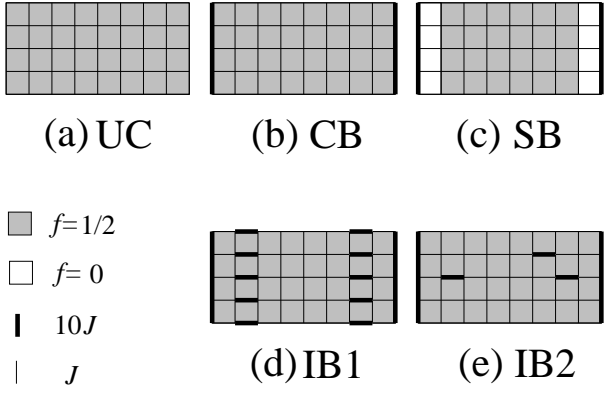}}
\vskip 0.5cm
\caption{ Current injection methods:
(a) Uniform current (UC) injection method, (b) conventional busbar (CB),
(c) busbar method in Ref.~\protect\onlinecite{simkin98} where
unfrustrated columns ($f=0$) are used near busbars (SB). 
(d) and (e) show busbar methods with inhomogeneities
near boundaries (IB1 and IB2, respectively). The fat links have a 
much larger coupling strength than the normal links.
}
\label{fig_models}
\end{figure}

We numerically study the fully frustrated $L_x \times L_y$ RSJ square arrays at zero temperature.
We use the periodic boundary condition in the $y$ direction and the
external current $I_d$ is inserted on the left boundary ($x=0$) and the extracted
on the right boundary ($x=L_x$). 
The current $I_{ij}$ from site $i$ to site $j$ is the sum of the supercurrent
and the normal current:
\[
I_{ij} = J_{ij} \sin(\theta_i - \theta_j + A_{ij}) + \dot{\theta_i} - \dot{\theta_j},
\]
where $J_{ij}$ is the critical current of the single junction, the
fully frustrated case corresponds to the magnetic bond angle 
$A_{ij}=\pi x$ for all vertical links at $x$
and zero for all links in the $x$ direction, and finally,  
$\dot{\theta_i} - \dot{\theta_j}$ is the normal current (in a
convenient unit system). The RSJ equations are obtained by
demanding current conservation on each site~\cite{khlee} and 
are integrated either by starting from the
ground-state configuration found from the simulated annealing
Monte Carlo method for $I_d=0$ or from a random configuration. 
The current injection methods are shown in Fig.~\ref{fig_models} and
are as follows: 
The uniform-current (UC) injection method~\cite{khlee}
corresponds to $J_{ij}=1$ for all links [Fig.~\ref{fig_models}(a)]; 
the conventional busbar (CB) method~\cite{simkin97} where the links on
the boundaries have large $J_{ij}$ corresponding to a
superconducting busbar (we use $J_{ij}=10$ for the fat links in 
Fig.~\ref{fig_models}); Simkin's improved busbar (SB) method~\cite{simkin98} 
where in addition to the superconducting
busbar there are no magnetic fields on the plaquettes adjacent to the boundaries
[Fig.~\ref{fig_models}(c)]; the busbar method with
inhomogeneities near the boundaries (IB) [Figs.~\ref{fig_models}(d) and (e) 
are two variants, IB1 and IB2, respectively]. The idea behind these
inhomogeneities close to the contacts is that
they create barriers for the onset of the vortex motion. 
The fat horizontal links in
Figs. 1(d) and (e) play the role of such barriers.   

\begin{figure}
\centerline{\epsfxsize=8cm \epsfbox{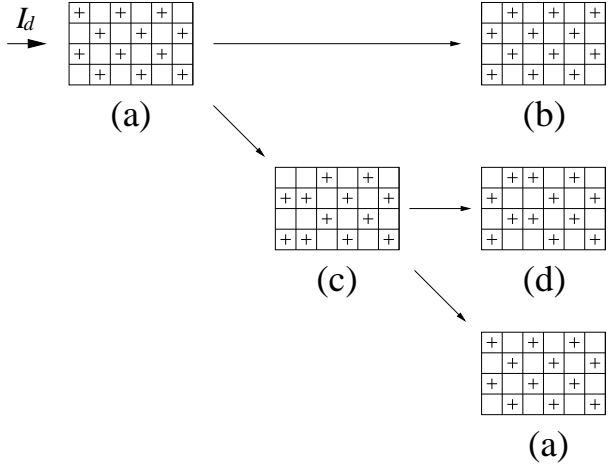}}
\vskip 0.5cm
\caption{ The configuration
with translational symmetry in (a) can evolve in different ways:
From (a) to (b), the translational symmetry is preserved by the rigid
motion of vortices.  From (a) to (c), only one column near boundary moves.
Once the motion (a) to (c) happens, this unstable vortex
configuration can be resolved by the motion either to (d) or to (a).
}

\label{fig_motion}
\end{figure}

The translationally invariant ground state is a checker board pattern
with a vortex at every second plaquette [see Fig.~\ref{fig_motion}(a)]. 
Let us first assume the translational symmetry even when an external current 
$I_d$ is turned on. Every second horizontal junction
will then by symmetry have the same current and we below focus on the
horizontal junctions with the largest current. For some external
current $I_d=I_s$ the current through these links reaches the maximum
supercurrent for a single junction, which is 1 in our current unit. 
A straightforward calculation shows
that this occurs at $I_s=(3-\sqrt{5})/2$ (see Table~\ref{tab_1}). For $I_d<I_s$
the ground state vortex configuration is always stable. However, at
some $I_d=I_c\geq I_s$ this ground state configuration becomes unstable
and dissipation sets in. In the translationally invariant case this
happens at $I_c=\sqrt{2}-1$ (see Table~\ref{tab_1}),~\cite{benz} which can
readily be verified with the boundary condition in Ref.~\onlinecite{bjkim-s}
which preserves the translational invariance; $I_c=0.4142(1)$ has been
obtained in excellent agreement with the prediction $\sqrt{2}-1$.

In reality, however, the current pattern enforced by the contacts is
not fully compatible with the translationally invariant vortex
pattern. One consequence of this is that the external current $I_{sb}$ needed
before a horizontal junction at the boundary reaches its maximum
supercurrent is lower than for the translationally invariant
case. Table~\ref{tab_1} gives the actual values for the boundary conditions
shown in Fig.~\ref{fig_models} together with the vortex column for which it occurs. This
lower value of $I_{sb}$ suggests that the vortex columns close to the
contacts may start to move and dissipate energy before the rest of the
vortices. Figure~\ref{fig_motion} illustrates some of the possibilities: The transition
from Fig.~\ref{fig_motion}(a) to ~\ref{fig_motion}(b) is the translationally 
invariant case where the
whole vortex lattice moves in one piece. The transition from Fig.~\ref{fig_motion}(a)
to \ref{fig_motion}(c) followed by \ref{fig_motion}(c) to \ref{fig_motion}(a) 
illustrates the case when only the vortex
column closest to the boundary moves and the others remain
fixed. Note that this is a boundary effect in the sense that it only
produces a voltage drop for the horizontal junctions closest to the
contacts and no voltage drop across the rest of the sample. Another
possibility is that the motion of the first vortex column 
[Fig.~\ref{fig_motion}(a) to ~\ref{fig_motion}(c)] 
is followed by the second column [as illustrated by Fig.~\ref{fig_motion}(c) 
to ~\ref{fig_motion}(d)], and then followed by the third and so on without stopping. Such a
consecutive motion of vortex columns starting from the contacts and
continuing without stopping we refer to as a domino-type motion.
We have found that this domino-type motion starts at a lower
current than what is required for the translationally invariant
motion and that consequently the threshold
current for the domino motion is the true critical current. Thus
the domino effect is responsible for the critical current for all
the boundary conditions shown in Fig.~\ref{fig_models} and the corresponding values
are given in Table~\ref{tab_1}.

The domino instability has two basic ingredients: The first is that the
boundary and current injection break the translationally invariant
current pattern close to the boundary which lowers the threshold
current for the first vortex column to move. This is related to that
$I_{s1}$ for the first column is smaller than $I_s$ for the
translationally invariant case (see Table~\ref{tab_1}). The second ingredient is
that this motion creates an instability which propagates through the
whole sample. A hand-waving argument for this goes as follows: One
first notes that as long as the checkerboard pattern of vortices is
present a horizontal junction with lower current is always followed
by one of higher current. However, when the first column moves
[as illustrated by Fig.~\ref{fig_motion}(a) to ~\ref{fig_motion}(c)] then two vortices become
neighbors. This means that now two horizontal junctions with higher
currents follow each other. Because of current conservation this means
that the junctions in the second column with higher currents will
momentarily further increase which makes this column unstable. The
motion of the second column in the same fashion causes the third
column to become unstable and so on.

\begin{figure}
\centerline{\epsfxsize=12cm \epsfbox[50 50 554 770] {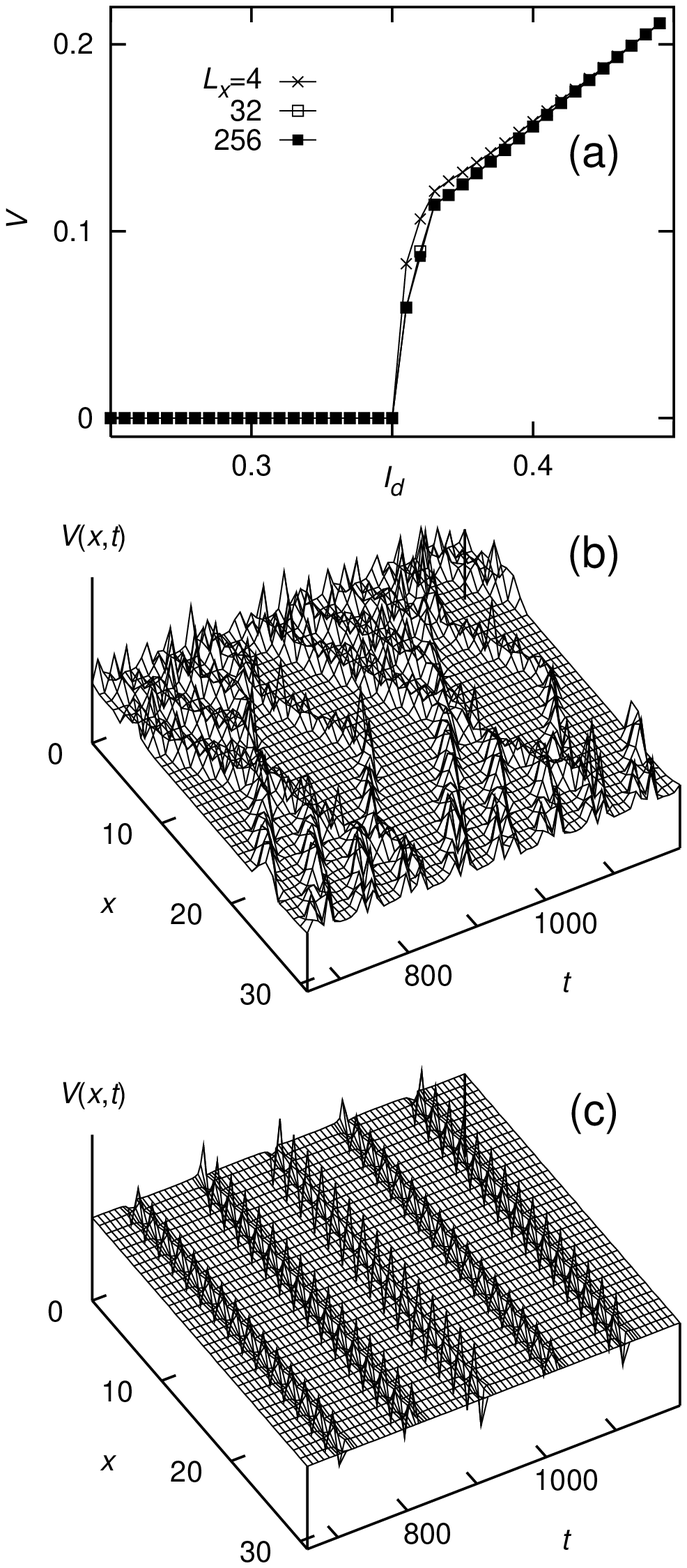}}
\caption{ (a) Time averaged voltage $V$ per junction versus external current $I_d$ for
uniform-current (UC) injection case. The absence of finite-size effects for
large sizes shows that $I_c = 0.350(1)$ is the true critical current.
(b) Voltage $V(x,t)$ at time $t$ across the horizontal junction at position $x$ 
along a row of junctions for UC
at $I_d = 0.36$: voltage pulses travel from the contacts into the sample.
(c) $V(x,t)$ for the translationally invariant case at $I_d = 0.417$: 
the voltage occurs instantaneously across the whole sample corresponding
to the rigid motion.
}
\label{fig_wave1}
\end{figure}

Figures~\ref{fig_wave1}(a) and (b) illustrate the domino mechanism for the uniform current
injection: Figure~\ref{fig_wave1}(a) shows that the voltage is zero up to a critical
current $I_c$. The fact that the voltage per length saturates to a
nonzero value above $I_c$ as the distance between the contacts
is increased, demonstrates that it is a true onset of dissipation
across the whole sample. Figure~\ref{fig_wave1}(b) shows the voltage drop $V(x,t)$
at time $t$ for the horizontal junction at position $x$ along a row of junctions 
in the presence of a current slightly larger than the critical current 
[$I_d = 0.36 > I_c = 0.350(1)$]. 
As seen the voltage pulse always travels
from one of the contacts towards the opposite in accordance with the
domino mechanism. In addition, the interference between domino propagations
traveling in opposite directions creates interesting pattern in this
3D plot. A measurable consequence of this is that, if the voltage drop
across different positions along the current direction is measured as
a function of time, then traveling voltage pulses can be observed. 
This is in contrast to the translationally invariant case
where the voltage appears at the same time along the whole current
direction as is illustrated in Fig.~\ref{fig_wave1}(c). This
translationally invariant case is simulated with the
fluctuating-twist-boundary method, as is described in Ref.~\onlinecite{bjkim-s}. 

\begin{figure}
\centerline{\epsfxsize=8cm \epsfbox[75 266 478 650]{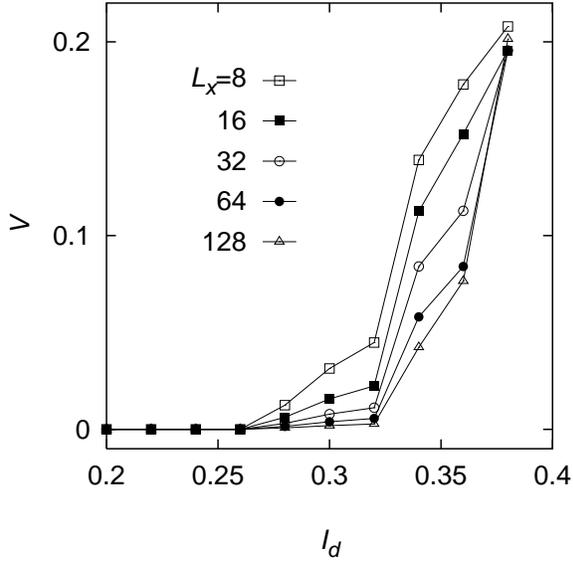}}
\vskip 0.5cm
\caption{Time averaged voltage $V$ per junction versus external current $I_d$ for
for the conventional busbar case [Fig.~\ref{fig_models}(b)]: 
$V$ between the boundary critical current $I_{sb}\approx 0.28$ and 
the true critical current $I_c\approx 0.32$ vanishes as the distance 
between the contacts increases because it is due to the motion of 
the vortex columns closest to the contact. 
The onset of nonzero voltage across the sample requires the domino mechanism 
which sets in at $I_c \approx 0.32$.
}

\label{fig_ivcb}
\end{figure}

The domino mechanism applies to all the situations shown in Fig.~\ref{fig_models}
with some variants: For example, in the case of the conventional busbar
injection [see Fig.~\ref{fig_models}(b)] the first vortex column becomes unstable at a
current $I_{cb}$ but the domino avalanche does not set in until a
larger current $I_c$ is applied (see Table~\ref{tab_1}). This is
illustrated in Fig.~\ref{fig_ivcb}, which shows the onset of dissipation at $I_{sb}$
for a finite system and a further increase at $I_c$. However, as the
distance between the contacts becomes larger dissipation between
$I_{cb}$ and $I_c$ decreases towards zero. The situation between
$I_{cb}$ and $I_c$ is hence a boundary effect of the type illustrated by 
Fig.~\ref{fig_motion}(a) to ~\ref{fig_motion}(c)
followed by ~\ref{fig_motion}(c) to ~\ref{fig_motion}(a), i.e., 
only the first column moves.

\begin{figure}
\centerline{\epsfxsize=11cm \epsfbox[100 266 604 648]{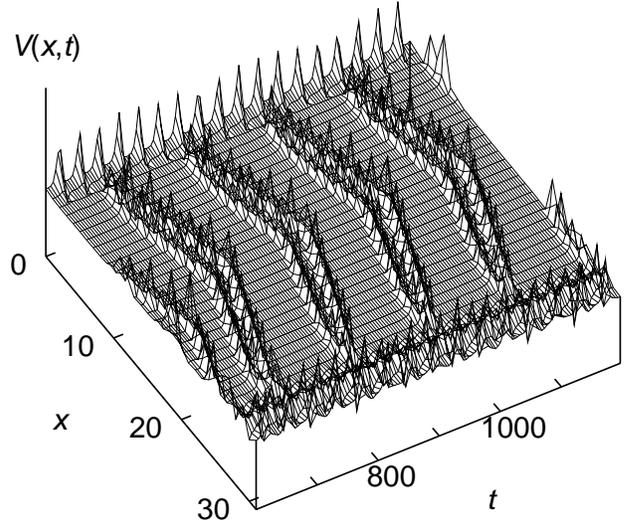}}
\caption{ Voltage $V(x,t)$ at time $t$ across the horizontal junction at 
position $x$ along a row of junctions for the IB2 case with
inhomogeneities close to the contacts in Fig.~\ref{fig_models}(e) at $I_d = 0.416$.
The critical current is in this case close to the upper limit given by the 
translationally invariant case, but the domino mechanism is still responsible
for the onset of nonzero voltage.
}
\label{fig_wave2}
\end{figure}

In Ref.~\onlinecite{benz} the critical current for a fully frustrated
square array of Josephson junctions was measured and found to be
$I_c\approx 0.42(2)$.~\cite{crit} As apparent from Table~\ref{tab_1} this value is larger
than what is obtained for the RSJ model with uniform-current injection 
[Fig.~\ref{fig_models}(a)] and busbar injections 
[Fig.~\ref{fig_models}(b) and (c)] and is in
fact close to the translationally invariant value. This means that the
domino mechanism is somehow suppressed. One way of suppressing
this mechanism is to introduce barriers close to the boundary which
prevents the domino avalanche from spreading. The situation in Fig.~\ref{fig_models}(d) shows
an extreme variant of this and as seen from Table~\ref{tab_1} this results in a
critical current close to the translationally invariant case. However,
also in this case the domino mechanism is responsible for the
onset of dissipation. Thus the translationally invariant value appears to be an upper
limit and the true critical current is always
lower and given by the domino mechanism. A perhaps more
realistic situation is with some imperfections close to the boundaries
which creates barriers for the vortex motion like in
Fig.~\ref{fig_models}(e)
and as seen from Table~\ref{tab_1} the critical current is again
close to the translationally invariant value. Figure~\ref{fig_wave2} shows the domino
mechanism for this case in a 3D plot. [Note the similarity with Fig.~\ref{fig_wave1}(b).]

The time scale of the Josephson junction array is given by
$t_0=\hbar/2eR_NJ$ with the shunt resistance $R_N$. For the
$1000\times1000$ $SNS$ junction array in Ref.~\onlinecite{benz} where
$R_N\approx 2$ ${\rm m}\Omega$ and $J\approx 7$ mA this means $t_0\approx
10^{-11}$ sec. The time for a voltage pulse to reach the middle would
then for the uniform current injection [Fig.~\ref{fig_wave1}(b)] as well as for the 
case with inhomogeneities
(Fig.~\ref{fig_wave2}) be of the order of  $10^{-7}$ sec. Thus it seems that it should
be possible to verify the domino mechanism by either direct voltage
measurements or by some vortex imaging technique.~\cite{lachenmann}

In summary, on the basis of simulations for the RSJ model, we have
concluded that the critical current for a 2D fully frustrated Josephson junction array is
due to a domino mechanism and that the voltage pulses resulting
from the domino avalanches should be open to experimental verification.

Discussions with C.~J. Lobb, R.~S. Newrock, and S. Teitel are gratefully acknowledged. 
The research was supported by the Swedish Natural Science Research Council
through Contract No. FU 04040-322.

\narrowtext

\begin{table}
\caption{ The critical current ($I_c$), the boundary critical current for
the motion of the first vortex column ($I_{cb}$), and the external current
needed for the first junction to reach its maximum current 
($I_{sb}$) (in $I_{si}$, $i$=1,2,5 denotes the column it occurs in). TI denotes
the translationally invariant case; see Fig.~\ref{fig_models} for 
the other current injection methods. }
\begin{tabular}{c c c c}
   &   $I_c$                       &  $I_{cb}$  &   $I_{sb}$  \\ \hline
TI &  $\sqrt{2} - 1 \approx 0.4142$& $ $           &  $I_{sb} = I_s = (3-\sqrt{5})/2 \approx 0.382 $\\
UC &  0.350(1)                     & $ $           &  $I_{s1} = 0.285(1) $ \\
CB &  0.324(1)                     & 0.277(1)   &  $I_{s1} = 0.183(1) $ \\
SB & 0.337(1)                      & $ $           &  $I_{s2} = 0.265(1) $ \\
IB1 & 0.4135(5)                    & $ $           &  $I_{s5} = 0.366(1) $ \\
IB2 & 0.4138(5)                    & 0.281(1)   &  $I_{s1} = 0.249(1) $ \\
\end{tabular}
\label{tab_1}
\end{table}

\end{multicols}


\begin{thebibliography}{99}
\bibitem{crit} This is estimated from the maximum of $dV/dI$ which
  occurs at a slightly higher value than the true threshold.
\end{thebibliography}
\end{document}